\documentclass[twocolumn,prb, showpacs]{revtex4-1}% Physical Review B
\usepackage{graphicx}% Include figure files
\usepackage{dcolumn}% Align table columns on decimal point
\usepackage{bm}% bold math
\usepackage{color}%Patrick's updates
\usepackage{ulem}%Patrick's updates

\begin{document}

%\preprint{}

\title{Oxygen non-stoichiometry and the origin of Na ion ordering in P2-Na$_{x}$CoO$_2$}

\author{G. J. Shu$^{1}$}
\author{W. L. Lee$^{2}$}
\author{F. -T. Huang$^{3,4}$}
\author{M. -W. Chu$^{1}$}
\author{Patrick A. Lee$^5$}
\author{F. C. Chou$^{1,6}$}
\email{fcchou@ntu.edu.tw}
\affiliation{
$^1$Center for Condensed Matter Sciences, National Taiwan University, Taipei 10617, Taiwan}
\affiliation{
$^2$Institute of Physics, Academia Sinica, Taipei 10115, Taiwan}
\affiliation{
$^3$Taiwan International Graduate Program, Academia Sinica, Taipei 10115, Taiwan}
\affiliation{
$^4$Department of Chemistry, National Taiwan University, Taipei 10617, Taiwan}
\affiliation{
$^5$Department of Physics, Massachusetts Institute of Technology, Cambridge, MA 02139, USA}
\affiliation{
$^6$National Synchrotron Radiation Research Center, Hsinchu 30076, Taiwan}

\date{\today}

\begin{abstract}
The impact of oxygen deficiency on physical properties of Na$_{2/3}$CoO$_{2-\delta}$ has been investigated.  From the combined thermogravimetric, magnetic susceptibility and synchrotron X-ray Laue diffraction studies, it is demonstrated that Na$_{2/3}$CoO$_2$ shows no superlattice ordering due to Na ions; however Na$_{2/3}$CoO$_{1.98}$, which has the same Co valence as that of Na$_{0.71}$CoO$_2$, shows nearly identical magnetic and transport properties and the same simple hexagonal superlattice ordering of $\sqrt{12}$a.  It is proposed that the Na ion ordering found in Na$_{2/3}$CoO$_{1.98}$ is identical to the ideal Na$_{0.71}$CoO$_2$ of large $\sqrt{12}$a$\times$$\sqrt{12}$a$\times$3c superlattice but with additional Na vacancies which are bound to the oxygen defects at room temperature.  We conclude that oxygen vacancies play a key role in stabilizing the superlattice structure and must be accounted for in its modeling.

\end{abstract}


\maketitle

\section{\label{sec:level1}Introduction\protect\\ }

P2-Na$_x$CoO$_2$ (also called $\gamma$-phase) is composed of alternating layers of Na and CoO$_2$ with a hexagonal in-plane network that can be described by space group P6$_3$/mmc.\cite{Fouassier1973}  Na vacancy ordering has been an important subject both experimentally and theoretically following the discovery of rich phase diagram of Na$_x$CoO$_2$,\cite{Foo2004} from the A-type antiferromagnetic (AF) ordering for x$\geq$0.75, to AF and metal-to-insulator transitions near x=0.5, and the superconductivity found in the hydrated sample of x $\sim$ 1/3.\cite{Takada2003}  In particular, the discovery of Na superlattice described by the multi-vacancy cluster model proposed by Roger \textit{et al.} has generated additional interest in the exploration of the origin of Na vacancy ordering and its impact on the title compound.\cite{Roger2007, Chou2008}  

We have demonstrated that the particularly stable phases found near 0.71, 0.50, 0.33, and 0.25 also show sharp drop of the Na atom diffusion coefficients at room temperature.\cite{Shu2008}  However, it is quite surprising that the commonly predicted stable phases near x = 3/4 and 2/3 do not exist, as revealed by our single crystal electrochemical de-intercalation studies at room temperature.  The Na superlattice as a result of Na ion ordering has been solved satisfactorily for x $\sim$ 0.833, 0.820, 0.77 and 0.71 so far, based on Na multi-vacancy model of di-, tri-, and quadri-vacancy cluster ordering.\cite{Chou2008, Shu2007, Shu2008, Huang2009, Shu2009}  For example, x = 0.71 is composed of alternating Na layers with tri- and quadri-vacancy clusters ordered in a large simple hexagonal superlattice of size $\sqrt{12}$a.\cite{Chou2008, Huang2009}  The implication of such Na vacancy ordering has helped us understand the origin of small cross sections of the Fermi surface hole pockets observed 
 by Shubnikov-de Haas oscillation and the intriguing coexistence of metallic conductivity with Curie-Weiss behavior. This led us to propose that the Na vacancy ordering creates one special partial localization environment for the doped holes, i.e., part of the charge is localized near the vacancy center and the rest are itinerant throughout the vacancy-free region.  

While Na vacancy is unavoidable for Na$_x$CoO$_2$ samples prepared under ambient pressure, even under pure oxygen environment, the most probable Na content for samples prepared using solid state reaction is usually in the range of 0.75-0.85.\cite{Motohashi2003}  Different vacancy ordering patterns for x = 2/3 have been proposed based on density functional theory calculations.\cite{Meng2008, Zhang2005}  However, single crystal diffraction studies on Na$_x$CoO$_2$ have not revealed any signature of Na ion ordering for x = 2/3,\cite{Shu2008}  On the other hand, Platova \textit{et al.} have proposed based on NMR and X-ray diffraction results, that a simple hexagonal superlattice of $\sqrt{12}$a exists in Na$_{2/3}$CoO$_2$ where powder sample were carefully prepared under oxygen environment at 700$^\circ$ C and quenched.\cite{Platova2009}  
The proposed microscopic structure is very different from the tri- and quadra-vacancy layer model we proposed for Na$_{0.71}$CoO$_2$.  It is very surprising that the physical properties of their samples are very similar to ours despite the difference in Na concentration and their 
proposed superlattice size is identical to that found in Na$_{0.71}$CoO$_2$ based synchrotron X-ray Laue and electron diffraction studies on single crystal sample with electrochemically fine-tuned x.\cite{Huang2009, Chou2008}  We have pointed out previously that the controversy could have resulted from differences in oxygen content,\cite{Huang2009} i.e., the identical $\sqrt{12}$a simple hexagonal superlattice is a result of identical average Co valence between Na$_{0.71}$CoO$_2$ and Na$_{2/3}$CoO$_{1.98}$.  Herein, we provide conclusive evidence to resolve the controversy regarding the $\sqrt{12}$a superlattice ordering observed in x = 2/3.  It turns out that oxygen content plays an important factor that cannot be ignored when discussing all related physical properties in the Na$_x$CoO$_{2-\delta}$ system.  The nature of immobile and randomly distributed oxygen vacancy at room temperature renders the Na ion ordering more complicated, especially when the ordering along the c-
 axis (staging) is considered as well.\cite{Shu2009, Huang2009}

\section{\label{sec:level1}Experiment\protect\\}

Single crystal samples have been prepared from floating zone method growth, following an additional electrochemical de-intercalation technique as reported previously.\cite{Shu2007}  Oxygen environment with pressure higher than 8 atm has been kept in the optical floating-zone growth chamber to reduce Na loss due to evaporation and the generation of Co$_3$O$_4$ impurity phase in particular, to prevent the occurrence of oxygen deficiency.  High quality single crystal Na$_x$CoO$_2$ of x $\sim$ 0.84 could be obtained reproducibly starting from polycrystalline feed rod prepared from Na$_2$CO$_3$ and Co$_3$O$_4$ of Na:Co=0.85:1 ratio.  Samples with Na content of 0.71 and 2/3 were prepared from electrochemical de-intercalation as reported previously,\cite{Shu2007} where the as-grown crystal sample of Na$_{0.84}$CoO$_2$ was used as working electrode, Pt as counter electrode, and Ag/AgCl with sieved liquid junction as reference electrode.  The applied voltages relative to Ag/AgCl refer
 ence electrode were 0.05V and 0.20V for the preparation of x=0.71 and 2/3, respectively.  Electron probe microanalysis (EPMA) has been performed on freshly cleaved inner surfaces to verify the chemical composition of each element.   Perkin-Elmer TG/DTA was used for thermal gravimetric analysis.  Synchrotron X-ray Laue diffraction for superstructure investigation was performed at the NSRRC, Taiwan and magnetic susceptibility measurement was done using Quantum Design SQUID-VSM.

\section{\label{sec:level1}Results and Discussions\protect\\ }

\begin{figure}
\begin{center}
\includegraphics[width=3.5in]{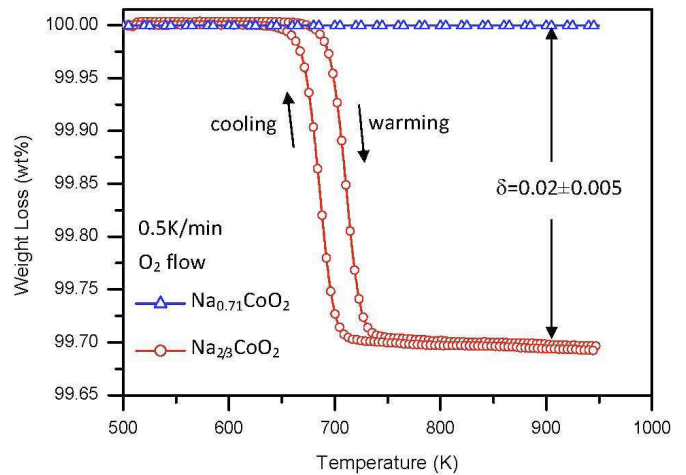}
\end{center}
\caption{\label{fig:fig-TGA}(Color online) TGA scan for Na$_{0.71}$CoO$_2$ and Na$_{2/3}$CoO$_2$ single crystal prepared by electrochemical technique. The scan has been done with warming/cooling rate of 0.5K/min in an oxygen environment.}
\end{figure}

Na$_{0.71}$CoO$_2$ sample can be prepared through electrochemical de-intercalation method precisely, as indicated by the diverging diffusion coefficient derived previously.\cite{Shu2008}  However, the target Na concentration for Na$_{2/3}$CoO$_2$ is very hard to obtain precisely, partly because of its metastable nature.  We have prepared sample with x = 0.67$\pm$0.01 and verified by both the X-ray lattice parameter analysis of linear c(x) extrapolation reported previously and EPMA to confirm this crucial concentration for this study.  Additional oxygen control has been performed using TGA scan under various gas flows, including oxygen and air, up to 950K.  The as-prepared single crystal samples of nearly oxygen deficiency-free samples for x = 0.71 and x = 2/3 have been ground into powder and go through additional annealing under TGA as shown in Fig.~\ref{fig:fig-TGA}.  We find that a significant weight loss occurs for x = 2/3 only near 700K and this weight loss is recoverable
  under oxygen.  On the other hand, x = 0.71 sample is stable at the same temperature range, whether the flowing gas is oxygen or air.  The reversible weight loss for x = 2/3 sample near 700K is about  0.30 wt$\%$, which corresponds to an oxygen loss of 0.02.  Interestingly, the Co valence of Na$_{0.71}$CoO$_2$ and Na$_{2/3}$CoO$_{1.98}$ are identical, and this thermal analysis suggests that the most stable average Co valence is near +3.29.  To prepare oxygen deficient Na$_{2/3}$CoO$_{1.98}$ single crystal sample for further study, we annealed the as-prepared Na$_{2/3}$CoO$_2$ single crystal sample at 700 $^\circ$C for 12 hours and quenched, similar to the method used by Platova \textit{et al.} on their powder sample.\cite{Platova2009}  We believe that the powder sample studied by Platova \textit{et al.} has an oxygen deficiency level about 0.02 as described above.

\begin{figure}
\begin{center}
\includegraphics[width=3.5in]{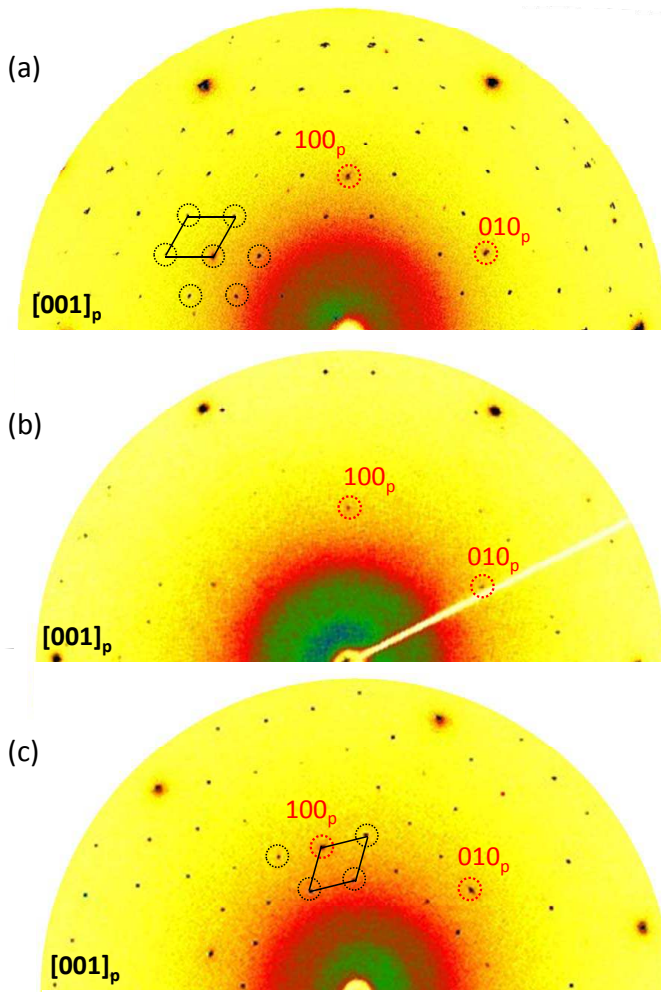}
\end{center}
\caption{\label{fig:fig-Laue}(Color online)  The Laue patterns for (a) Na$_{0.71}$CoO$_2$, (b) Na$_{2/3}$CoO$_2$, and (c) Na$_{2/3}$CoO$_{1.98}$ taken along [001] direction at room temperature. The strong diffraction peaks are indexed as [100] and [010] of the primitive cell with size \textbf{a} constructed by Co ions, and the six superlattice peaks in hexagonal shape surrounding each primitive index correspond to a real lattice of size $\sqrt{12}$a. }
\end{figure}

It is instructive to compare the Laue diffraction patterns of Na$_{0.71}$CoO$_2$, Na$_{2/3}$CoO$_2$, and Na$_{2/3}$CoO$_{1.98}$ as shown in Fig.~\ref{fig:fig-Laue}. The Laue pattern of Na$_{0.71}$CoO$_2$ shows perfect superlattice of $\sqrt{12}$a as reported previously.\cite{Chou2008}  The superstructure of Na$_{0.71}$CoO$_2$ has been modeled to be composed of alternating layers of sodium tri- (T) and quadri-vacancy (Q) clusters ordered in simple hexagonal shape.\cite{Chou2008}  Additional structure refinement suggests that these alternating layers of TQTQTQ stacking has a 3c periodicity described by a right-hand screw symmetry advanced Na-trimers along the c-axis.\cite{Huang2009}  While Na$_{2/3}$CoO$_2$ shows no superlattice formation, as indicated in Fig.~\ref{fig:fig-Laue}(b), Na$_{2/3}$CoO$_{1.98}$ reveals a superstructure nearly identical to that of Na$_{0.71}$CoO$_2$ in the reciprocal space.  We note however, that  detailed comparison of Laue patterns in Fig.~\ref{fig:fig-Laue} shows some systematic missing of superlattice diffraction planes for Na$_{2/3}$CoO$_{1.98}$ as comparing with that of Na$_{0.71}$CoO$_2$, which suggests that even though these two samples possess identical superlattice size of $\sqrt{12}$a, subtle difference must exists as suggested by the NMR studies.\cite{Alloul2009}  These results seem to suggest that similar Na ion ordering can be achieved as long as the average Co valence is close to +3.29.

Experimentally Na$_{0.71}$CoO$_2$ has been shown to be the most stable phase as indicated by its significant sharp drop of the diffusion coefficient and the convincingly refined crystal structure of Na tri- and quadri-vacancy cluster stacking of $\sqrt{12}$a$\times$$\sqrt{12}$a$\times$3c superlattice ordering,\cite{Chou2008, Shu2008, Huang2009} the observed identical superlattice size suggests that Na$_{2/3}$CoO$_{1.98}$ can very well be a defect type variation of the ideal Na$_{0.71}$CoO$_2$.  To provide a more quantitative description to the current finding, let us take as an empirical fact that Na$_{0.71}$CoO$_2$ with the tri- and quadri-vacancy ordering is most stable.  The Na$_{2/3}$CoO$_{1.98}$ with oxygen vacancy means that for every 48 Co atoms, one oxygen is missing,  i.e., 1 out of 4 $\sqrt{12}$a$\times$$\sqrt{12}$a supercell contains an oxygen vacancy.  In order not to pay too much Coulomb energy for the O$^{2-}$ vacancy two additional Na vacancies will appear.  Then on average the two tri-vacancy layers become quadra-vacancies resulting in 8 Na per 12 cobalt atoms, i.e., we are led to a defect 
type Na$_{0.71}$CoO$_2$ structure of Na$_{2/3}$CoO$_{1.98}$ stoichiometry.  Another way of saying is that the oxygen vacancy is frozen in place randomly but the Na is mobile at room temperature. The additional missing Na (relative to 0.71) will be bound close to the missing oxygen in order to maintain the $\sqrt{12}$a superlattice.  The reason for this stability presumably has large contribution from Madelung energy, so it is at least partly electron driven.  In fact Nicolaou \textit{et al.} have also proposed that the stacking layer potential could create a special charge ordering within the CoO$_2$ layer based on their ARPES experimental results.\cite{Nicolaou2010}  

The superlattice found in Na$_{2/3}$CoO$_{1.98}$ must be robust because it is prepared from a high temperature quenching procedure of easy disorder introduction, while Na$_{2/3}$CoO$_2$ is prepared from a slow cooling process under oxygen atmosphere.  We must note that oxygen vacancy is not mobile at room temperature while Na ion self diffusion is active above $\sim$150-200K.  The randomly distributed oxygen vacancy at the level of $\sim$0.02 per formula unit for Na$_{2/3}$CoO$_{1.98}$ must be tolerated in order to allow Na layer to form a superstructure similar to that of Na$_{0.71}$CoO$_2$, It is difficult to understand within the proposal of Platova \textit{et al}. for Na$_{2/3}$CoO$_2$ why oxygen vacancies are needed to stabilize the superlattice structure.  In any case, based on our findings it is clear that the role of oxygen deficiencies need to be accounted for in the modeling of the $x = 2/3$ material.

\begin{figure}
\begin{center}
\includegraphics[width=2.5in, angle= -90]{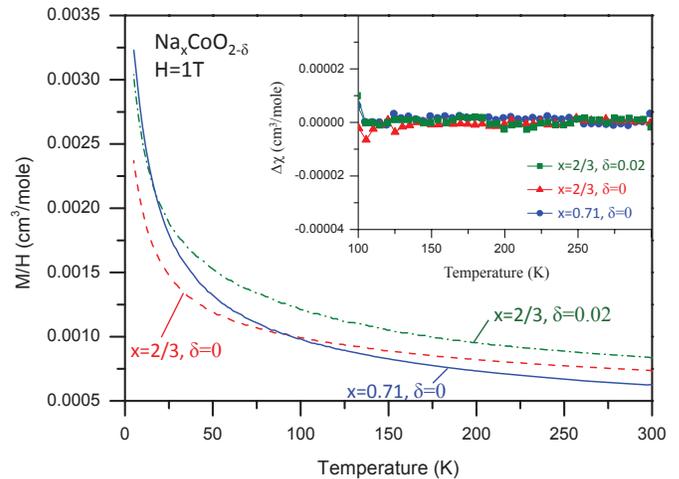}
\end{center}
\caption{\label{fig:fig-chi}(Color online) Magnetic susceptibility data for Na$_{0.71}$CoO$_2$, Na$_{2/3}$CoO$_2$, and Na$_{2/3}$CoO$_{1.98}$ in applied field of 1~Tesla.  The inset demonstrates the reliability of Curie-Weiss law fitting through difference between measured data and fitted equation of $\chi_\circ$+C/(T-$\theta$) with parameters shown in Table~\ref{tab:tableI}.}
\end{figure}

%Note: {table} is single column and {table*} is double column
\begin{table}
\begin{center}
\caption{\label{tab:tableI}Summary of magnetic susceptibility data analysis for Na$_{2/3}$CoO$_2$, Na$_{2/3}$CoO$_{1.98}$, and Na$_{0.71}$CoO$_2$ fitted using equation $\chi_\circ$+C/(T-$\theta$) in the range of $\sim$100-300K. }
\begin{tabular}{cccccc}
\hline\hline
 sample & $\chi_\circ$(cm$^3$/mole) & C(cm$^3$ K/mole) & $\theta$ ) \\
\hline
Na$_{2/3}$CoO$_2$                         &0.00049(3)    & 0.099(2)    & -96(2) \\
\hline
Na$_{2/3}$CoO$_{1.98}$                 &0.00054(5)    & 0.109(2)    & -60(2) \\
\hline
Na$_{0.71}$CoO$_{2}$                     & 0.00034(2)    &0.106(1)   & -62(1) \\
\hline\hline
\end{tabular}
\end{center}
\end{table}

Magnetic susceptibility measurement results are summarized in Fig.~\ref{fig:fig-chi}.  The displayed data are powder average of the single crystal measurement, i.e., 2/3$\chi_{ab}$+1/3$\chi_c$ without considering the minor correction of in-plane anisotropy.  There is no magnetic anomaly found down to 2K for all three samples.  The Curie-Weiss behavior for a system with metallic property has long been a puzzle,\cite{Foo2004} and it transforms from Curie-Weiss to Pauli-like behavior only when x is reduced to the level of nearly half-filled Na layer.\cite{Shu2007}  Based on our proposed Na vacancy cluster model published previously,\cite{Chou2008} Na tri- and quadri-vacancy cluster superlattice ordering could be responsible for the Curie-Weiss metal behavior, i.e., part of the doped holes are localized near the Na vacancy cluster center while the rest are itinerant.  Both localized and itinerant electrons should contribute to the magnetic susceptibility.  We propose that the total magnetic susceptibility should include contributions from temperature independent core diamagnetic $\chi^{core}$, van Vleck paramagnetic $\chi^{VV}$, Pauli paramagnetic $\chi^{Pauli}$, and the spin susceptibility from localized spins which are tentatively described by a Curie-Weiss like behavior of C/(T-$\theta$). The Curie-Weiss behavior for all three samples has been fitted following an equation of $\chi_\circ$+C/(T-$\theta$) in the range of 100-300K quite satisfactorily as seen in the inset of Fig.~\ref{fig:fig-chi}, where C corresponds to the number of localized spins, Weiss temperature $\theta$ indicates the mean field average of the magnetic coupling strength among isolated spins, and $\chi_\circ$ = $\chi^{core}$+$\chi^{VV}$+$\chi^{Pauli}$.  The fitted parameters for $\chi$(T) described by the Curie-Weiss law are shown in Table~\ref{tab:tableI}.  The fitted $\chi_\circ$ values do not provide reliable physical meaning before each individual contribution can be analyzed.  It is not surprising to find that C and $\theta$ are about the same for Na$_{0.71}$CoO$_2$ and Na$_{2/3}$CoO$_{1.98}$, which implies that the quantity and the coupling strength of isolated spins are nearly identical.  Na$_{2/3}$CoO$_2$ shows less isolated spins even though more holes are introduced compared with Na$_{2/3}$CoO$_{1.98}$.  Clearly, formation of long-range Na multi-vacancy cluster ordering such as that demonstrated by the ideal Na$_{0.71}$CoO$_2$, is prohibited in Na$_{2/3}$CoO$_2$, unless oxygen vacancy is introduced.

The co-existence of local moments and itinerant electrons is a complicated many-body problem which is not fully understood.  The data fitting using Curie-Weiss law plus Pauli is only a crude beginning.  For instance, the $C$ coefficient found for Na$_{0.71}$CoO$_2$ corresponds to a $S = {1\over 2}$ local moment per doped hole, which is clearly inconsistent with metallic behavior.  We also acknowledge that another point of view has been proposed in the literature, i.e., the Curie-Weiss behavior signifies an almost ferromagnetic metal\cite{Alloul08}, as described, for instance, by the self consistent renormalized (SCR) spin fluctuation theory.\cite{Moriya} However, in the standard SCR theory, the spin susceptibility saturates at low temperatures leading to Fermi liquid behavior whereas in the present system $\chi$ continues to rise down to 4~K (as seen in Fig.3).  Such a low Fermi liquid scale is not expected without fine tuning.  Furthermore, the resistivity exhibits a linear $T$ dependence below 100~K and cross-over to $T^{1.3}$ below 4~K, while the specific heat exhibits $T \ln T$ dependence.\cite{Balicas2008}  At the moment, neither the itinerant spin fluctuation model nor the partial local moment picture is capable of explaining these anomalies.  We leave this complicated and interesting issue open for future discussions.  The main point of the present analysis is simply that the magnetic properties of Na$_{0.71}$CoO$_2$ and Na$_{2/3}$CoO$_{1.98}$ are very similar, while that of Na$_{2/3}$CoO$_2$ has distinctly different $C$ and $\theta$.

\begin{figure}
\begin{center}
\includegraphics[width=2.5in, angle= -90]{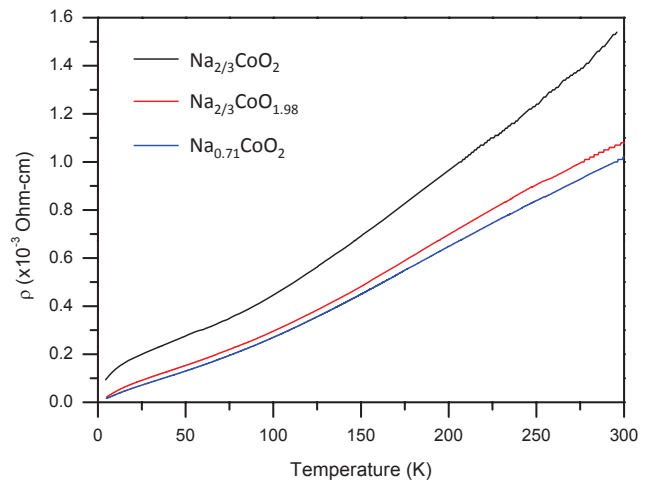}
\end{center}
\caption{\label{fig:fig-resistivity}(Color online) In-plane resistivities for Na$_{0.71}$CoO$_2$, Na$_{2/3}$CoO$_2$, and Na$_{2/3}$CoO$_{1.98}$.}
\end{figure}

\begin{figure}
\begin{center}
\includegraphics[width=3.5in, angle= -90]{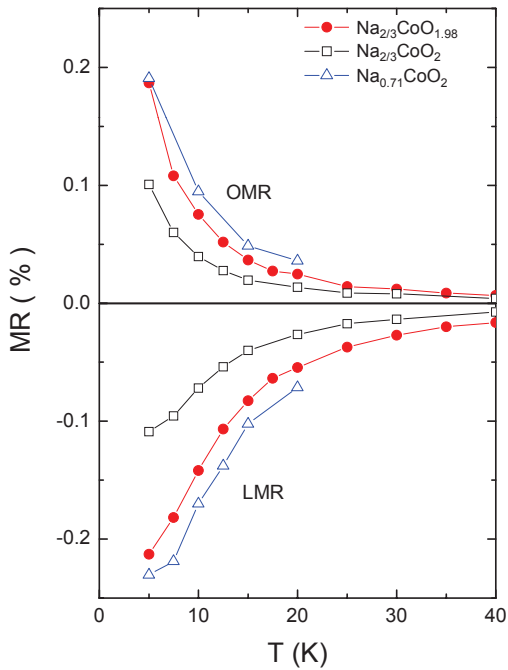}
\end{center}
\caption{\label{fig:fig-MR}(Color online) Longitudinal (LMR) and orbital (OMR) magnetoresistance for Na$_{0.71}$CoO$_2$, Na$_{2/3}$CoO$_2$, and Na$_{2/3}$CoO$_{1.98}$ are compared under magnetic field of 14 Tesla. }
\end{figure}

We have also performed 4-probe resistivity measurements on the Na$_{0.71}$CoO$_2$, Na$_{2/3}$CoO$_2$, and Na$_{2/3}$CoO$_{1.98}$ crystals as shown in Fig~\ref{fig:fig-resistivity}. For Na$_{0.71}$CoO$_2$, the resistivity shows metallic behavior and drops with decreasing temperature. There is a crossover at about 100~K, below which the resistivity becomes practically linear in T.  Similar behavior has also been observed in Na$_{2/3}$CoO$_2$ and Na$_{2/3}$CoO$_{1.98}$.  However, such deviation from the T-linear dependence at low temperature is evident and more significant in Na$_{2/3}$CoO$_2$.  In addition, we note that the resistivity value is more than twice as large in Na$_{2/3}$CoO$_2$ at low temperature, which suggests a shorter carrier lifetime due to a more disordered phase in Na$_{2/3}$CoO$_2$.  These observations are in accord with the absence of Na superstructure ordering as mentioned earlier and also imply an intimate correlation between the Na ordering and the Co la
 yer electronic structure.            

Longitudinal magnetoresistance (LMR) and transverse magnetoresistance (TMR) were measured for the three samples, where LMR and TMR were obtained under a magnetic field up to 14 Tesla applied parallel and perpendicular to the CoO$_2$ plane, respectively.  The orbital magnetoresistance (OMR) was determined by subtracting TMR with LMR in order to exclude the spin related contribution to the TMR. A comparison of LMR and OMR for the three samples is shown in Fig.~\ref{fig:fig-MR}. The OMR is positive for all three samples with quadratic magnetic field dependence as expected from orbital contribution, which is simply proportional to the square of carrier mobility.  Na$_{2/3}$CoO$_2$ shows smaller OMR than that of Na$_{0.71}$CoO$_2$ and Na$_{2/3}$CoO$_{1.98}$, which infers a shorter carrier lifetime in a more disordered Na$_{2/3}$CoO$_2$ as expected.  On the other hand, LMR is typically derived from spin effect, where the negative LMR is most likely associated with the suppression of spin scattering due to the field alignment of localized moment.  We find that the LMR in Na$_{2/3}$CoO$_2$ is nearly half of that in Na$_{0.71}$CoO$_2$ and Na$_{2/3}$CoO$_{1.98}$ at low temperature, which suggests less spin scatterers in the system.  Again, the main point is the similarity between Na$_{0.71}$CoO$_2$ and Na$_{2/3}$CoO$_{1.98}$ and their contrast with Na$_{2/3}$CoO$_2$.
We also remark that the extraction of carrier density from Hall coefficient data is not possible because of its strong temperature dependence due to strongly correlated nature.

%\subsection{\label{sec:level2}Magnetic susceptibilities\protect\\}

%\begin{eqnarray*}
%\chi &=  \chi_{_{0}} + N_{chain} \chi_{_{chain}} + N_{_{D}} \chi_{_{Dimer}} + N_{_{CW}} \chi_{_{CW}}, \\
    % &=  \chi_{_{0}} + N_{chain} \frac{1}{T} \sum_{n,k} a_{n,k} \alpha^k (\frac{J_{_{1}}}{k_{_{B}}T})^n + \\
     %&  N_{D}\frac{N_A g^2 \mu_{_B}^2}{k_{_{B}}T (3 + e^{\frac{J_{_{D}}}{k_{_{B}}T}})} + N_{_{CW}} \frac{C}{T-\Theta}.
%\label{eq:Buhler}
%\end{eqnarray*}
%\noindent The only constraint used in the fitting process is the normalization of N$_{chain}$+N$_{dimer}$+N

\section{\label{sec:level1}Conclusions\protect\\ }

  In summary, using single crystal samples of well controlled oxygen and sodium content, we have provided strong evidence to resolve the controversy of superlattice ordering found in Na$_x$CoO$_2$ of x=0.71 and 2/3.  Synchrotron X-ray Laue diffraction results indicate that there is no sodium ordering in Na$_{2/3}$CoO$_2$ at room temperature, while Na$_{2/3}$CoO$_{1.98}$, which has identical Co valence to that of Na$_{0.71}$CoO$_2$, shows identical $\sqrt{12}$a simple hexagonal superstructure to that of Na$_{0.71}$CoO$_2$, also with nearly identical magnetic and transport properties. These results imply that Na$_{0.71}$CoO$_2$ with $\sqrt{12}$a$\times$$\sqrt{12}$a$\times$3c superlattice is the most stable structure at this particular concentration, while Na$_{2/3}$CoO$_{1.98}$ of identical average Co valence possesses a nearly identical Na superstructure and magnetic properties with the help of oxygen defect generation.  This proposed model is supported in part by
  the higher resistivity 
%and deviation from the T-linear behavior at low temperature 
found in Na$_{2/3}$CoO$_2$ without Na superstructure.  Studies on Na ion ordering based on screened inter-sodium interaction without considering the involved oxygen deficiency could result in misleading prediction or interpretation.

\section{Acknowledgment}
GJS and FCC acknowledge the support from National Science Council of Taiwan under project number NSC-98-2119-M-002-021. We also thank Chung-Yuan Kao in NTU for the help on EPMA measurement. PAL acknowledges the support of DOE grant DE-FG02-03ER46076. \\
\appendix

%\bibliography{Na0.67-reference}

\end{document}